\documentclass[aps,prl,reprint,aps,preprintnumbers,showpacs]{revtex4-2}
\usepackage{amssymb,amsmath}
\usepackage{tensor}
\usepackage{hyperref}
\hypersetup{
	breaklinks=true,   
	colorlinks=false,   
    pdfusetitle=true,  
        }

\newcommand{\form}[1]{\mathrm{d}#1}

\newcommand{\scom}[2]{\left[#1\stackrel{\star}{,}#2\right]}

\begin{document}

\title{Quadratic twist-noncommutative gauge theory}
\author{Tim Meier}
\email{tmeier@physik.hu-berlin.de}
\author{Stijn J. van Tongeren}
\email{svantongeren@physik.hu-berlin.de}
\affiliation{Institut f\"ur Physik, Humboldt-Universit\"at zu Berlin, IRIS Geb\"aude, Zum Grossen Windkanal 2, 12489 Berlin}

\begin{abstract}
Studies of noncommutative gauge theory have mainly focused on noncommutative spacetimes with constant noncommutative structure, with little known about actions for noncommutative 4D Yang-Mills theory beyond this case. We construct an action for Yang-Mills theory on a quadratically-noncommutative spacetime, i.e. of quantum-plane type, obtained from a Drinfeld twist, with star-gauge symmetry. Applied to supersymmetric Yang-Mills theory, this gives a candidate AdS/CFT dual of string theory on a related deformation of AdS$_5\times$S$^5$, which is expected to be integrable in the planar limit.
\end{abstract}


\preprint{HU-EP-23/03-RTG}

\maketitle

Noncommutativity between space-time coordinates is a likely feature of quantum gravity \cite{Doplicher:1994tu}, actively studied from numerous angles \cite{Arzano:2021scz,Addazi:2021xuf}. In string theory, noncommutative gauge theory appears in the low energy dynamics of open strings \cite{Seiberg:1999vs}, and thereby the AdS/CFT correspondence \cite{Maldacena:1997re,Maldacena:1999mh,Hashimoto:1999ut}. Despite the formal and phenomenological relevance of noncommutative gauge theory, it is not clear how to write actions to all orders in the noncommutativity, when going beyond the case of constant noncommutativity. In this letter we consider a noncommutative Drinfeld-twist deformation of Minkowski space, with quadratically-coordinate-dependent noncommutativity, and construct an all-order action for Yang-Mills theory with star-gauge symmetry. Beyond providing a first example of a noncommutative Yang-Mills theory action with quadratic noncommutativity, our choice of deformation is motivated by the AdS/CFT correspondence and integrability. Applied to maximally supersymmetric Yang-Mills theory, our noncommutative deformation provides a concrete candidate gauge theory dual of a particular Yang-Baxter deformation \cite{Delduc:2013qra,Kawaguchi:2014qwa,vanTongeren:2015soa} of the famously integrable AdS$_5\times$S$^5$ superstring \cite{Arutyunov:2009ga,Beisert:2010jr}, as conjectured in \cite{vanTongeren:2015uha}. This opens the door to investigating integrability for a range of novel planar noncommutative gauge theories.

We consider noncommutative field theory in the usual spirit of Weyl quantization, trading noncommuting field operators for a noncommutative product -- the star product -- between commutative fields \cite{Madore:2000en,Szabo:2001kg}. Our star product is obtained from a Drinfeld twist, whereby it automatically comes with clear algebraic properties and a natural differential calculus \cite{Aschieri:2009zz}, and twists rather than plainly breaks Poincar\'e symmetry \cite{Chaichian:2004za,Wess:2003da}. The original Groenewold-Moyal noncommutative deformation can be viewed as a twist, and it is well known how to construct a Yang-Mills action in this case \cite{Szabo:2001kg}. For general twists, however, it is not clear how to define a suitable dual field strength tensor, and construct an action for noncommutative Yang-Mills theory. For example, the approaches of \cite{Dimitrijevic:2011jg,Dimitrijevic:2014dxa} for $\kappa$-Minkowski space, were necessarily perturbative, and only solved to leading order in the noncommutativity. To our knowledge, the only non-constant case known to all orders, is the $U(1)$ Yang-Mills theory studied in \cite{Ciric:2017rnf} for a particular twist with linear noncommutativity, where standard Hodge duality suffices.

We show how a twisted version of Hodge duality allows us to define a noncommutative Yang-Mills action for a twist based on two commuting Lorentz generators, with a noncommutative structure with quadratic coordinate dependence. Our construction moreover provides a broader framework that covers all known Poincar\'e-based twist-deformations of Minkowski space, including non-abelian ones, whose $r$ matrices are unimodular \cite{unimodularpoincaretwistpaper}. Given our motivations in AdS/CFT, we also discuss how to couple our theory to (adjoint) matter, define our deformed maximally-supersymmetric Yang-Mills theory, and discuss its AdS/CFT interpretation. 

\section{Lorentz-deformed Minkowski space}

In deformation quantization, a noncommutative spacetime is described via a regular spacetime whose function algebra is equipped with a noncommutative (star) product. The noncommutative product we consider for functions on $\mathbb{R}^{1,3}$ is based on the Drinfeld twist
\begin{equation}
\label{eq:Lorentztwist}
\mathcal{F} = e^{\frac{i\lambda}{2}(M_{01} \otimes M_{23} - M_{23} \otimes M_{01})},
\end{equation}
built from two commuting Lorentz generators, $M_{01}$ and $M_{23}$, with $M_{\mu\nu} = i (x_\mu \partial_\nu - x_\nu \partial_\mu)$. It defines our noncommutative star product via
\begin{equation}
\label{eq:starproductdef}
f(x)\star g(x) \equiv \mu(\mathcal{F}^{-1} (f(x),g(x))),
\end{equation}
where $\mu(h(x),z(x))=h(x)z(x)$ is the ordinary pointwise product of functions, and $\lambda$ is our deformation parameter. We will refer to this as the Lorentz twist, or deformation. In contrast to the familiar Groenewold-Moyal star product associated to
\begin{equation*}
\mathcal{F}_{GM} = e^{-\frac{i\theta^{\mu\nu}}{4}(\partial_\mu \otimes \partial_\nu - \partial_\nu \otimes \partial_\mu)},
\end{equation*}
where $\theta$ is a constant antisymmetric matrix, the bidifferential operator appearing in the exponent of our twist has nontrivial coordinate dependence \footnote{Changing coordinates does not fundamentally change this, as e.g. in polar coordinates in the $(2,3)$ plane, while $M_{23}$ becomes $\partial_\theta$, we should consider $e^{i\theta}$ rather than $\theta$ as an operator before the Weyl map, and $M_{23}(e^{i \theta})$ is not constant. Moreover, the natural Rindler type coordinates for $M_{01}$, inconveniently do not cover $\mathbb{R}^{1,3}$ in one go.}.

Our star product is associative because $M_{01}$ and $M_{23}$ commute \footnote{All twist-star products are associative thanks to the cocycle condition on the twist. We are purposefully introducing only minimal formal structure.}. Its noncommutative structure is
\begin{equation}
\label{eq:quadraticallorder}
x^\mu \star x^\nu = R_\sigma{}^{\mu}{}_\rho{}^{\nu} x^\rho \star x^\sigma ,
\end{equation}
with $24$ nonzero components for $R$,
\begin{align}
R_0{}^0{}_0{}^0=1, \quad R_0{}^0{}_2{}^2=\cosh{\lambda}, \quad R_0{}^1{}_2{}^3=-i\sinh{\lambda},
\end{align}
and others obtained by index permutation symmetries of the twist \eqref{eq:Lorentztwist}, namely $2\leftrightarrow3$ or $(0,1) \leftrightarrow(2,3)$ combined with a sign change of $\lambda$, and $0\leftrightarrow1$.

As our goal is gauge theory, we need differential calculus, now suitably twisted \cite{Aschieri:2009zz}. Using standard differential calculus on Minkowski space we define
\begin{equation}
\label{eq:formstarfunctiondef}
\mathrm{d}x^\mu \star f = \mu(\mathcal{F}^{-1}(dx^\mu,f)),
\end{equation}
where the vector fields $M_{01}$ and $M_{23}$ act via Lie derivatives, summarized as
\begin{equation}
\label{eq:Fbarmunudef}
\mathrm{d}x^\mu\star f   = \mathrm{d}x^\nu \bar{F}{}_\nu{}^\mu(f),
\end{equation}
with
\begin{equation*}
\bar{F}_\mu{}^\nu = \tensor{{\small \begin{pmatrix}
\cosh\frac{\lambda M_{23}}{2} & -\sinh\frac{\lambda M_{23}}{2}&0&0\\
-\sinh\frac{\lambda M_{23}}{2} & \cosh\frac{\lambda M_{23}}{2} &0 &0\\
0 & 0 & \cos\frac{\lambda M_{01}}{2} & -\sin\frac{\lambda M_{01}}{2}\\
0 & 0 &  \sin\frac{\lambda M_{01}}{2} & \cos\frac{\lambda M_{01}}{2}
\end{pmatrix}}}{_\mu^\nu}.\\
\end{equation*}
Commuting functions through forms gives rise to 
\begin{equation}
\label{eq:dxfComRule}
\form{x^\mu}\star f=R_\nu{}^\mu(f)\star\form{x^\nu},
\end{equation}
with the $R$ matrix $R_\nu{}^\mu = \bar{F}_\nu{}^\rho \bar{F}_\rho{}^\nu$. Both $R$ and $\bar{F}$ are vector-field-valued elements of the Lorentz group, in the sense that, raising and lowering indices with the usual Minkowski metric,
\begin{equation}
\label{eq:RandFinLorentzgroup}
R_\mu{}^\nu R^\rho{}_\nu = \bar{F}_\mu{}^\nu \bar{F}^\rho{}_\nu =\delta_\mu^\rho.
\end{equation}
We also define a star-wedge product
\begin{equation}
\mathrm{d}x^\mu \wedge_\star \mathrm{d}x^\nu = \hat{\mu}(\mathcal{F}^{-1}(\mathrm{d}x^\mu,\mathrm{d}x^\nu)),
\end{equation}
where $\hat{\mu}(a,b) = a \wedge b$ is the regular wedge product. Concretely
\begin{equation}
\label{eq:Fbar4indexdef}
\mathrm{d}x^\mu \wedge_\star \mathrm{d}x^\nu  = \bar{F}_\sigma{}^\mu{}_\rho{}^\nu \mathrm{d}x^\sigma \wedge \mathrm{d}x^\rho ,
\end{equation}
with $\bar{F}_\sigma{}^\mu{}_\rho{}^\nu=R_\sigma{}^\mu{}_\rho{}^\nu|_{\lambda \rightarrow \lambda/2}$. We will mostly work with star forms
\begin{equation}
\omega = \omega^\star_{\mu\nu\ldots \rho} \star \mathrm{d}x^\mu \wedge_\star \mathrm{d}x^\nu \wedge_\star\ldots \wedge_\star \mathrm{d}x^\rho,
\end{equation}
but occasionally will also express them as regular forms
\begin{equation}
\omega = \omega_{\mu\nu\ldots \rho} \mathrm{d}x^\mu \wedge \mathrm{d}x^\nu \wedge\ldots \wedge \mathrm{d}x^\rho.
\end{equation}
Our star forms are totally R-antisymmetric, e.g.
\begin{equation*}
\mathrm{d}x^\mu \wedge_\star \mathrm{d}x^\nu  = -  R_\rho{}^\mu{}_\sigma{}^\nu \mathrm{d}x^\sigma \wedge_\star \mathrm{d}x^\rho.
\end{equation*}
We will use the ordinary exterior derivative, which has the desired product rule
\begin{equation}
\mathrm{d}(\omega \wedge_\star \chi) = \mathrm{d}\omega \wedge_\star \chi + (-1)^p \omega \wedge_\star d\chi.
\end{equation}
for $p$ and $q$ forms $\omega$ and $\chi$, respectively, as it commutes with Lie derivatives. Under (conventional) conjugation we have $\overline{\omega \wedge_\star \chi} = (-1)^{pq} \overline{\chi} \wedge_\star \overline{\omega}$.

Our star product is graded cyclic under an integral,
\begin{equation}
\int\omega\wedge_\star\chi= (-1)^{p}\int \chi\wedge_\star\omega
\end{equation}
when $\chi\wedge_\star\omega$ is a top form, upon integration by parts \cite{Aschieri:2009ky}.

\section{Hodge duality}

To define our twisted Hodge star we take a natural generalization of the Levi-Civita symbol,
\begin{align}
\label{eq:generalizedLeviCivita}
\form{x^\mu}\wedge_\star\form{x^\nu}\wedge_\star\form{x^\rho}\wedge_\star\form{x^\sigma}=\epsilon^{\mu\nu\rho\sigma}\form{{}^4x},
\end{align}
where the volume form $\form{{}^4x} =\form{x^0} \wedge_\star \form{x^1} \wedge_\star \form{x^2} \wedge_\star \form{x^3} = \form{x^0} \wedge\form{x^1} \wedge\form{x^2} \wedge \form{x^3}$ is not deformed. 
This $\epsilon$ is graded cyclic, and its $32$ nonzero components are
\begin{equation}
\begin{aligned}
\epsilon^{0123} & = -\epsilon^{0132}=\epsilon^{0231}=-\epsilon^{0321} =1,\\
\epsilon^{1212} & = -\epsilon^{0202}=\epsilon^{1313}= -\epsilon^{0303}=i \sinh{\lambda},\\
\epsilon^{0312} & = -\epsilon^{0213}=\cosh{\lambda},
\end{aligned}
\end{equation}
plus others related by graded cyclicity.

In regular Hodge duality we can freely permute indices on the Levi-Civita symbol for signs, giving many equivalent definitions of a dual form. The appropriate choice in our twisted setting is
\begin{widetext}
\begin{equation}
* \mathrm{d}x^{\mu_1} \wedge_\star \ldots \wedge_\star \mathrm{d}x^{\mu_k} = \tfrac{(-1)^{\sigma(k)}}{(4-k)!} \epsilon_{\mu_{k+1} \ldots \mu_{4}}{}^{\mu_1\ldots \mu_k} \mathrm{d}x^{\mu_4} \wedge_\star \ldots \wedge_\star \mathrm{d}x^{\mu_{k+1}},
\end{equation}
\end{widetext}
where $\sigma(p)$ denotes the signature of the reversal of $p$ objects, i.e. $\sigma(1)=\sigma(4)=0$, $\sigma(2)=\sigma(3)=1$. The reversed index contraction in the dual form is essential. Restricted to basis star forms, this twisted Hodge star commutes with Lie derivatives along vector fields in the Poincar\'e algebra, and hence with our star product. This allows us to consistently extend it star-linearly to arbitrary forms, where it continues to commute with Poincar\'e Lie derivatives and our star product.

Our Hodge star has all typical properties, appropriately twisted \cite{unimodularpoincaretwistpaper}. It preserves R antisymmetry and reality of star forms, and for a $p$ form $\omega$ we have
\begin{equation}
** \omega = -(-1)^p \omega.
\end{equation}
For equal-degree $p$ forms $\omega$ and $\chi$ we also have
\begin{align*}
\omega \wedge_\star *\chi  & = (-1)^p * \omega \wedge_\star \chi \\
& = (-1)^{\sigma(p)+1} p! \omega^\star_{\mu \ldots \nu} \star R_{\kappa}{}^\mu \ldots R_{\rho}{}^\nu \chi^{\star \rho \ldots \kappa} \,\form{{}^4x},
\end{align*}
so that
\begin{equation}
\int \omega \wedge_\star *\chi = \int \chi \wedge_\star * \omega.
\end{equation}
Related to Hodge duality commuting with star products,
\begin{equation}
\scom{\form{x^\mu} \wedge_\star \form{x^\nu} \wedge_\star \form{x^\rho} \wedge_\star \form{x^\sigma}}{f}=0,
\end{equation}
for any $f$, following from star-commutativity of $\epsilon$ and the volume form, these being constant and Lorentz invariant respectively. This implies that $\epsilon$ is an invariant of the $R$ matrix,
\begin{equation}
\label{eq:epsilonRinvariant}
\epsilon^{\tau\kappa\zeta\phi}\tensor{R}{_\tau^\mu}\tensor{R}{_\kappa^\nu}\tensor{R}{_\zeta^\rho}\tensor{R}{_\phi^\sigma} = \epsilon^{\mu\nu\rho\sigma}.
\end{equation}
A similar form of Hodge duality was discussed for $q$-Minkowski space in \cite{Meyer:1994wi}, see also \cite{Majid:1994mh}.

\section{Yang-Mills theory}

Coming to Yang-Mills theory, as gauge transformations are functions, they are affected by the star product, and it is natural to consider star-gauge transformations \cite{Szabo:2001kg}. A fundamental field $\Phi$ then transforms as
\begin{align*}
\delta_\varepsilon\Phi(x)&=i\varepsilon(x)\star\Phi(x),
\end{align*}
under a gauge transformation by $\varepsilon \in \mathfrak{h}$, where $\mathfrak{h}$ is the Lie algebra of the gauge group $H$\footnote{Star-gauge transformations only close for transformations in the fundamental representation of $\mathfrak{u}(n)$ (or $\mathfrak{gl}(n,\mathbb{C})$, undesirable for other reasons). Other algebras can be considered in a universal enveloping algebra approach \cite{Jurco:2001rq}, with the infinitely many associated degrees of freedom reduced to finitely many via the Seiberg-Witten map \cite{Seiberg:1999vs}, at least perturbatively in the deformation parameter. For our AdS/CFT applications, $\mathfrak{u}(n)$ suffices.}.

Working in terms of forms, we have
\begin{align*}
\mathrm{d}\left(\delta_\varepsilon\Phi\right)=\mathrm{d}\left(i\varepsilon\star\Phi\right)=i\mathrm{d}\varepsilon\star\Phi+i\varepsilon\star\mathrm{d}\Phi,
\end{align*}
and we can define the covariant derivative
\begin{equation}
\label{eq:covDerPhi}
\mathrm{D}\Phi=\mathrm{d}\Phi+iA\star\Phi,
\end{equation}
with
\begin{equation*}
\delta_\varepsilon A=\mathrm{d} \varepsilon+i\scom{\varepsilon}{A},\qquad \delta_\varepsilon\left(\mathrm{D}\Phi\right)=i \varepsilon \star \mathrm{D}\Phi.
\end{equation*}

Next we define the field strength tensor
\begin{equation}
G=\mathrm{d}A-iA\wedge_\star A,
\end{equation}
which transforms star-covariantly
\begin{equation*}
\delta_\varepsilon G=i\scom{\varepsilon}{G}.
\end{equation*}
We now consider a natural deformation of the commutative Yang-Mills action,
\begin{equation}
\label{eq:YMAction}
S_{\text{NC-YM}}=\int\mbox{Tr}~ G\wedge_\star *G.
\end{equation}
Since our Hodge dual commutes with star products, $*G$ transforms star covariantly. Since our star product is cyclic under integration, this action is gauge invariant.

To illustrate this nontrivial point, let us derive the transformation of $*G$ in components. Starting from
\begin{equation*}
G=G^\star_{\mu\nu} \star \form{x^\mu} \wedge_\star \form{x^\nu},
\end{equation*}
using eqs. \eqref{eq:dxfComRule}, we find
\begin{equation}
\delta_\varepsilon G^\star_{\mu\nu}=i\varepsilon \star G^\star_{\mu\nu}-i G^\star_{\rho\sigma}\star \tensor{R}{_\mu^\rho}\tensor{R}{_\nu^\sigma} \varepsilon.
\end{equation}
The transformation of $*G$ is then
\begin{align}
\delta_\varepsilon\left(*G\right)& =i\varepsilon \star (*G)\notag \\
& \quad -i G^\star_{\mu\nu}\epsilon_{\xi\kappa}{}^{\tau\lambda} \star \form{x^\rho} \wedge_\star \form{x^\sigma} \star \tensor{R}{^\xi_\sigma}\tensor{R}{^\kappa_\rho}\tensor{R}{_\tau^\mu}\tensor{R}{_\lambda^\nu}\varepsilon\notag \\
& = i \scom{\varepsilon}{*G},
\end{align}
where we used eqs. \eqref{eq:RandFinLorentzgroup} and \eqref{eq:epsilonRinvariant}.

In star components our action reads
\begin{equation}
\label{eq:YMstarcomponents}
S_{\text{NC-YM}} = \int\mbox{Tr}~ G^\star_{\mu\nu} \star R_\rho{}^\mu R_\sigma{}^\nu G^{\star\sigma\rho} \form{{}^4x}.
\end{equation}
Expressed in unstarred components, repeated integration by parts gives
\begin{equation}
\label{eq:YMnonstarcomponents}
S_{\text{NC-YM}} = \int\mbox{Tr}~ G_{\mu\nu}  G^{\nu\mu} \,\form{{}^4x},
\end{equation}
where
\begin{equation}
\label{eq:Gnonstarcomponents}
G_{\mu\nu} = \partial_{[\mu} A_{\nu]} - i \bar{F}_\rho{}^\kappa{}_\sigma{}^\tau \bar{F}_{[\nu|}{}^\sigma (A_\kappa) \star \bar{F}^\rho{}_{|\mu]} (A_\tau),
\end{equation}
showing that the kinetic term for the gauge field is undeformed, while the interaction terms are deformed.

Our action has twisted Poincar\'e symmetry in the spirit of \cite{Chaichian:2004za,Wess:2003da}, meaning the following. In the commutative setting, the Poincar\'e algebra acts on individual fields via Lie derivatives, which by the product rule combine to Lie derivatives of the Lagrangian \footnote{In our covariant notation, this relies on the fact that Poincar\'e Lie derivatives commute with Hodge duality. The same is required, and holds, in our twisted setting.}. For Poincar\'e generators these are total derivatives, leaving the action invariant. Introducing a coproduct $\Delta(\xi) = \xi \otimes 1 + 1 \otimes \xi$ for generators $\xi$, the product rule takes the form
\begin{equation*}
\xi(\mu(f,g)) =\mu(\Delta(\xi)(f,g)),
\end{equation*}
with multiple coproducts extending this to products involving more fields. Our twisted product is similarly compatible with a twisted coproduct
\begin{equation*}
\xi(f \star g) = \xi(\mu(\mathcal{F}^{-1}(f,g)))= \mu(\mathcal{F}^{-1} \Delta_\mathcal{F}(\xi)(f,g)),
\end{equation*}
where $\Delta_\mathcal{F} = \mathcal{F} \Delta \mathcal{F}^{-1}$. Since every product in our action is a star product, letting the Poincar\'e algebra act (nonlocally) on products of fields by this twisted coproduct, still result in a total derivative, and an invariant action.
The twisted Poincar\'e algebra for our Lorentz twist is discussed in \cite{Lukierski:2005fc}.

\section{Matter fields and supersymmetric Yang-Mills theory}

We can readily couple our theory to matter. For adjoint scalars for instance, we can write
\begin{equation}
S_{\text{NC-}\phi}=\int\mbox{Tr}~\mathrm{D}\phi^\dagger\wedge_\star*\mathrm{D}\phi + \int \mbox{Tr}(\phi^\dagger \star \phi)^{\star n}\, \form{{}^4x}.
\end{equation}
where $\mathrm{D}\phi=\form{\phi}-i\scom{A}{\phi}$. Gauge invariance follows as for star-Yang-Mills theory.

Working with forms allows us to straightforwardly define actions, while guessing e.g. the component forms of eqs.  (\ref{eq:YMstarcomponents}-\ref{eq:Gnonstarcomponents}) would be difficult. To tackle fermions in similar spirit, we combine left and right-handed Weyl spinors $\psi_\alpha$ and $\bar{\psi}_{\dot{\alpha}}$ with Grassmann-valued basis spinors $s^\alpha$ and $\bar{s}^{\dot{\alpha}}$ to form the Grassmann-even $\psi =\psi_\alpha s^\alpha$ and $\bar{\psi} = \bar{\psi}_{\dot{\alpha}} s^{\dot{\alpha}}$. We then take our twist to act via the left and right-handed Weyl representation of the Poincar\'e algebra, on $s^\alpha$ and $\bar{s}^{\dot{\alpha}}$ respectively. These spinors play an analogous role to forms, in components resulting in spinor analogues of eqs. (\ref{eq:formstarfunctiondef}-\ref{eq:Fbar4indexdef}).


We now assemble our $\gamma$ matrices into a convenient object, taking the Pauli matrices $\sigma_{i}$, $i=1,2,3$, and $\sigma_0 = 1_{2\times2}$ to form
\begin{align*}
\sigma&=\tensor{\sigma}{_\mu_{\alpha\dot{\alpha}}}s^\alpha\bar{s}^{\dot{\alpha}}\form{x^\mu}=\tensor{\sigma}{^\star_\mu_{\alpha\dot{\alpha}}}s^\alpha \star \bar{s}^{\dot{\alpha}} \star \form{x^\mu}.
\end{align*}
Coupled by the Pauli matrices, the transformation properties of the spinors and one form cancel, making $\sigma$ Lorentz invariant, hence star commutative. For adjoint fermions we then define the kinetic action
\begin{align}
S_{\text{NC-}\psi} = \int\int\int\mathrm{d}^2s\mathrm{d}^2\bar{s}~\mbox{Tr}~ \bar{\psi} \star \sigma \wedge_\star *\mathrm{D}\psi,
\end{align}
where $\mathrm{D}\psi=\form{\psi}-i\scom{A}{\psi}$, and the Grassmann integrals over the basis spinors extract the appropriate components. Gauge invariance of this action follows as before, since $\sigma$ is star commutative. Combined with an adjoint scalar $\phi$, we can form gauge-invariant Yukawa-like interactions such as
\begin{equation*}
\int\int\mathrm{d}^2s~\mbox{Tr}~\psi \star \phi \star \psi~ \mathrm{d}^4x.
\end{equation*}

We use these ingredients to define the action for maximally-supersymmetric Yang-Mills theory (SYM) on Lorentz-deformed $\mathbb{R}^{1,3}$ as
\begin{widetext}
\begin{align}
S_{\text{NC-SYM}}=& \frac{1}{4g^2}\mbox{Tr}\int G\wedge_\star *G + \mbox{Tr}\int \mathrm{D}\phi^{IJ} \wedge_\star *\mathrm{D}\phi_{IJ} -\frac{g^2}{16}\mbox{Tr}\int\mathrm{d}^4x~ \scom{\phi^{IJ}}{\phi^{KL}}\star\scom{\phi_{IJ}}{\phi_{KL}}\\
&\notag +\mbox{Tr}\int\mathrm{d}^2s\mathrm{d}^2\bar{s}\int \bar{\psi}^I\star\sigma\wedge_\star*\mathrm{D}\psi_I +\frac{ig}{2}\mbox{Tr}\int\mathrm{d}^2s\int\mathrm{d}^4x~\psi_I\star\scom{\phi^{IJ}}{\psi_J}-\frac{ig}{2}\mbox{Tr}\int\mathrm{d}^2\bar{s}\int\mathrm{d}^4x~\bar{\psi}^I\star\scom{\phi_{IJ}}{\bar{\psi}^J}.
\end{align}
\end{widetext}
where $\psi^I$, $I=1,2,3,4$, are the four fermions of SYM, and the $\phi^{IJ}=-\phi^{JI}$ contain the six real scalars. This deformation of SYM classically has twisted superconformal symmetry \cite{unimodularpoincaretwistpaper}. As the dilatation generator commutes with our twist, this action is conventionally scale invariant.

With gauge algebra $\mathfrak{u}(n)$ this action provides a candidate AdS/CFT dual to the Yang-Baxter deformation of the AdS$_5\times$S$^5$ superstring \cite{Delduc:2013qra,Kawaguchi:2014qwa,vanTongeren:2015soa} for the $r$ matrix $r = M_{01} \otimes M_{23} - M_{23} \otimes M_{01}$, as conjectured in \cite{vanTongeren:2015uha}, see also \cite{vanTongeren:2016eeb}, based on a shared twisted symmetry structure, and conceptually in line with the discussion in \cite{Beisert:2005if}. The corresponding AdS$_5$ background is deformed to
\begin{align}
ds^2 & = \frac{-\rho^2 \mathrm{d}\alpha^2 +r^2 \mathrm{d}\theta^2}{z^2-\tilde{\lambda}^2 \rho^2 r^2 /z^2} +\frac{\mathrm{d}\theta^2 + \mathrm{d}r^2+\mathrm{d}z^2}{z^2},\\
B & = \tilde{\lambda} \frac{\rho^2 r^2}{z^4-\tilde{\lambda}^2 \rho^2 r^2} \mathrm{d}\alpha \wedge \mathrm{d}\theta, \quad e^{-2(\phi-\phi_0)} =1-\lambda^2 \frac{\rho^2 r^2}{z^4},\notag
\end{align}
in Rindler coordinates $(\rho,\alpha)$ in the $(x^0,x^1)$ plane, and polar coordinates $(r,\theta)$ in the $(x^2,x^3)$ plane, of AdS$_5$ in the Poincar\'e patch. It is further supported by nontrivial Ramond-Ramond forms. The deformation parameters are related as $\lambda = \sqrt{g_{\mbox{\tiny YM}}^2 N_c} \tilde{\lambda} / 2\pi$.

\section{Outlook}

We have constructed an action for noncommutative Yang-Mills theory with star-gauge symmetry, for the Lorentz twist with quadratic noncommutativity. Our construction relies on properties of the twist and R matrix, combined with our nontrivial twisted Hodge duality, and, for SYM, on our fermionic extension of twisted differential calculus. 


There are various open questions surrounding our deformation at the quantum level, for instance regarding UV/IR mixing and its presumable absence in SYM, and the fate of twisted symmetry. At the classical level, noncommutative gauge theories admit an underlying $L_\infty$ algebraic structure \cite{Blumenhagen:2018kwq,Giotopoulos:2021ieg}, and it would be interesting to investigate this for our deformation, and contrast it with the braided noncommutative gauge theories of \cite{Ciric:2021rhi,Giotopoulos:2021ieg}.

Applied to SYM, the Lorentz deformation gives the natural AdS/CFT dual of a related Yang-Baxter deformation of the AdS$_5$ string. Our construction in fact extends to all noncommutative spacetimes described by the known Drinfeld twists of the Poincar\'e algebra, with unimodular $r$ matrix, providing candidate gauge theory duals for a large class of Yang-Baxter deformations of the AdS$_5$ string \cite{unimodularpoincaretwistpaper}. However, while matching planar symmetry structures between gauge and string theory is certainly promising, the actual decoupling limit underlying these dualities can be subtle. For constant noncommutativity, while the spacelike and lightlike cases are fine \cite{Aharony:2000gz}, timelike noncommutativity results in a noncommutative open string, rather than gauge theory \cite{Seiberg:2000ms,Gopakumar:2000na}. The Lorentz deformation mixes these cases, appearing spacelike inside the lightcone in the $(0,1)$ plane, but timelike outside it, and in general this decoupling limit needs careful analysis. However, even in cases with a subtle decoupling limit, remnants of a duality to our type of noncommutative gauge theory are likely to survive at the planar level.

We expect planar Lorentz-deformed SYM to be integrable, based on the integrability of its proposed string dual. At the classical level this should take the form of Yangian invariance \cite{Beisert:2017pnr,Beisert:2018zxs}, now twisted similarly to \cite{Garus:2017bgl}. At the quantum level, we should find a spectral problem described by an integrable spin chain, similar to the famous dilatation operator of undeformed SYM \cite{Beisert:2003jj}. The Lorentz deformation is particularly natural in this regard, as it preserves dilatation symmetry. We have defined a suitable related spectral problem in planar Lorentz-deformed SYM, and are in the process of extracting its integrable structure \cite{NCintegrabilitypaper} -- which we expect to relate to the twisted spin chain of \cite{Beisert:2005if} -- building on a planar equivalence theorem \cite{unimodularpoincaretwistpaper} in the spirit of Filk \cite{Filk:1996dm}.  We hope this will pave the way to integrable AdS/CFT for general (homogeneous) Yang-Baxter deformations of the AdS$_5$ string, and its lower dimensional cousins.

\section{Acknowledgements.} We would like to thank Riccardo Borsato, Ben Hoare, and Anna Pacho\l{} for discussions, and Gleb Arutyunov, Riccardo Borsato, Jerzy Lukierski, Anna Pacho\l{}, and Richard Szabo for valuable comments on the draft. TM's research is funded by the Deutsche Forschungsgemeinschaft (DFG, German Research Foundation) - Projektnummer 417533893/GRK2575 ``Rethinking Quantum Field Theory''. The work of ST is supported by the German Research Foundation via the Emmy Noether program ``Exact Results in Extended Holography''. ST is supported by LT.

\end{document}